\documentclass[preprint,aps,showpacs, floatfix]{revtex4}

\usepackage{amsmath}
\usepackage{graphicx}
\usepackage{ifsym}
\usepackage{dcolumn}

\begin{document}
\title{Spatial Distribution of Gaussian Fluctuations
of the Molecular Field and Magnetization 
in the Pyramid-like Ising Nanoscopic System
Interacting with the Substrate}

\author{L.S. Borkowski and Z. Jacyna-Onyszkiewicz}
\author{}
\affiliation{Quantum Physics Division, Faculty of Physics,\\
A. Mickiewicz University, Umultowska 85, 61-614 Poznan, Poland}

\begin{abstract}
We study thermodynamic properties of an Ising model of a ferromagnetic
nanoscopic pyramid deposited onto a ferromagnetic bulk substrate.
The influence of the interaction between the pyramid and the substrate 
is calculated in terms of the equilibrium reduced-state (density) operator
used for description of thermodynamic properties of nanoscopic systems.
The spatial distribution of the fluctuations of molecular field 
and magnetization in the nanoscopic
pyramid is obtained in the Gaussian fluctuations approximation.
Experimental consequences for the magnetic force measurements
are briefly discussed.
\end{abstract}

\pacs{75.75.+a}

\maketitle

\renewcommand{\thesection}{\arabic{section}}
\section{Introduction}

During the past two decades we have witnessed significant
advances in the ability to synthesize nanoscale structures
as well as development of novel experimental method
allowing exploration of their physical properties.\cite{bland}
This is exciting for two reasons.
Firstly, new forms of matter with no counterpart in nature
and revealing unique physical properties have been fabricated.
Secondly, we have now
realized nanostructures that open new avenues
for development of very small devices.

Nanoscopic magnetic systems\cite{skomski2003} have always been very attractive
from the theoretical point of view.
On the other hand, studies of nanoscopic systems are characterized
by a close coupling between theory and experiment because of rapidly
increasing number of experimental works
on real nanoscopic materials.\cite{bland}
These materials often correspond remarkably close to certain idealized spin
models and are great challenges not only for physicists, but also
for chemists and engineers.

There is a great interest in producing two-dimensional arrays
of magnetic nanodots which may serve e.g. as magnetic recording media.
Due to small dimensions of these particles quantum-mechanical
effects cannot be ignored. Effects observed in quantum dots
are relevant also in quantum computing and spin electronics.
It was shown recently that nanoscopic structures in the form
of piramids or similar shapes can be fabricated
on a bulk substrate.\cite{nanogov2000,giersig, giersig2}

In experiments aiming at the investigation of nanoscopic systems
we deal frequently with such systems deposited on a bulk substrate.
However, most often in the theoretical description
of such experiments the interaction between the nanosystem
and the substrate is neglected. The main purpose of this work
is to find a way to include this and calculate
its effect on the thermodynamic properties of the Ising model
of a ferromagnetic nanoscopic system. To achieve this
we apply the equilibrium reduced-state (density) operator (ERSO)\cite{zjo02}
and carry out the calculations within the Gaussian fluctuations
approximation (GFA).

ERSO is the most general equilibrium state operator. It has been derived
from the generalized Schr\"odinger variational principle 
and not on the quantum statistical mechanics, i.e. without statistical
hypotheses.
When the term describing the interaction between the system
and its environment is neglected ERSO takes the form
of the statistical operator of the Gibbs canonical distribution.
ERSO applied to an exactly solvable microscopic model leads to exact
results.

GFA is a modified version of the high density expansion method as has
been proposed in refs. \cite{zo1980,zo1980A}. GFA is an improvement
over the molecular
field approximation (MFA) due to the self-consistent
inclusion of Gaussian fluctuations of this field. The essential
new element of GFA is the summing up of the partial sums
of Feynman diagrams of the same structure of recurrent formulae
at each stage of the calculations. GFA is based
on a classification of the Feynman diagrams in terms of $1/z$,
where $z$ is the effective number of spins interacting with
any given spin.
Owing to this procedure the theory becomes internally consistent
and does not lead to unphysical results such as, for example,
a complex Curie temperature.\cite{zo1986}

\section{The model}

In this paper we consider magnetic properties
of a nanoscopic pyramid (nanopyramid).
They are well described by the simple model of localized
and ordered spins with the following spin-1/2 Hamiltonian

\begin{equation}
\label{ham}
H = - \frac{1}{2} I \sum S^z_{{\bf f}l} S^z_{{\bf f^\prime}l^\prime} ,
\end{equation}

\noindent
where $I$ is the coupling parameter and $\sum$ stands for summation
over pairs of different simple cubic (sc) lattice points.
Our considerations are restricted to nearest-neighbor interaction only.

In expression (1) $\bf f$ denotes the two-dimensional position vectors
of a spin belonging to a given monoatomic layer $l=1, 2, 3, 4$
of the pyramid.
There are 4 spins in the $l=1$ layer, 16 spins in the $l=2$ layer,
36 spins in the $l=3$ layer and 64 spins in the $l=4$ layer.
The total number of spins in the pyramid is 120 (see Fig. 1).

We assume that the nanoscopic pyramid is deposited on a bulk ferromagnetic
substrate sufficiently well described by the spin-$1/2$ Ising model
for a simple cubic lattice and the Hamiltonian

\begin{equation}
\label{ham2}
H_s = -\frac{1}{2} I_1 \sum S^z_{{\bf g}r} S^z_{{\bf g^\prime}r^\prime} .
\end{equation}

Here ${\bf g}$ denotes the two-dimensional position vectors of spin
belonging to a given monoatomic layer. The summations always run
over different sites. The substrate is divided into monoatomic layers
parallel to the planes (100) of a sc-lattice. The position of each
layer is given by the number $r=1, 2, ...$

Let us assume that the interaction of the nanoscopic pyramid
with the bulk substrate is described
by the Heisenberg Hamiltonian of the form

\begin{equation}
\label{ham3}
H_I = - \frac{1}{2} I_2 \sum_{{\bf f}{\bf g}}
{\bf S}_{{\bf f}l=4} {\bf S}_{{\bf g}r=1} .
\end{equation}

In order to take into account the interaction of the nanoscopic
pyramid with the bulk substrate we shall apply ERSO\cite{zjo02}
suitable for description
of a physical situation similar to the one we are concerned with.
In the derivation of this particular form of ERSO
we use the fact that although
the Universe as a whole is in the pure state, its arbitrary multiparticle
parts are inevitably in mixed states. This is purely a quantum effect
following from the holistic properties of the quantum theory, formally
related to the fact that the Universe (according
to the quantum cosmology postulates) has one vector of state common for
all systems and, in the case of interactions among them,
we are not able to specify the vector of state
for individual subsystem (it may not be
the case for the interactions
of the effective field type or classical ones).
Such a situation does not occur in the classical description because we
may know classical trajectories of each particular molecule
irrespective of their interactions. The holistic features
of the quantum theory imply the use of the formalism
of ERSO in description of multiparticle systems. For this reason
the interaction (\ref{ham3}) of the nanoscopic
pyramid with the bulk substrate
cannot be described by the Ising Hamiltonian.

In our case ERSO
takes the following form\cite{zjo02}:

\begin{equation}
\label{rdo}
d = \exp [\beta (F - H - H^\prime(\beta))] ,
\end{equation}

\noindent
where $F$ is the free energy and

\begin{equation} 
\label{ham5}
H^\prime(\beta) = {\rm Tr}_s [H_I (1+K)\exp (\beta (F_s - H_s)) ]
\end{equation}

\noindent
is an effective term describing the interaction between the pyramid
and the bulk substrate, ${\rm Tr}_s [...]$ is the partial trace
over the substrate states,
$K$ is the correlation operator,
$\beta = (k_B T)^{-1}$,
and 
\begin{equation}
F_s = - \frac{1}{\beta} \ln {\rm Tr}_s [\exp (-\beta H_s)]
\end{equation}
is the free energy of the substrate.
Substituting expresssion (\ref{ham3}) into Eq. (\ref{ham5})
with the assumption that

\begin{equation}
| < S^x_{{\bf g}r} > |, |< S^y_{{\bf g}r} >|
\ll |< S^z_{{\bf g}r} >| \quad , \quad I_2 \ll I_1 \quad ,
\end{equation}
and $K\simeq 0$, we arrive at

\begin{equation}
H^\prime(\beta) = - \frac{1}{2} I_2 <S^z_{r=1}>
\sum_{\bf f} S^z_{{\bf f}l=4} \quad ,
\end{equation}
where 

\begin{equation}
\label{Szr1}
<S^z_{r=1}> = {\rm Tr}_s
[ S^z_{{\bf g}r=1} \exp (\beta (F_s - H_s))]
\end{equation}
is the average bulk substrate spin moment in the layer $r=1$.

\section{Gaussian fluctuation of the molecular field}

As a starting point to GFA we choose the following decomposition
of the Hamiltonian (\ref{ham})

\begin{equation}
H=(H-H_1)+H_1=H_0+H_1, 
\end{equation}
\noindent
where the perturbative part $H_1$ is defined by the tranformation

\begin{equation}
H \rightarrow H_1=H(S^z_{{\bf f}l} \rightarrow \delta S^z_{{\bf f}l}),
\end{equation}
\noindent
where

\begin{equation}
\delta S^z_{{\bf f}l} = S^z_{{\bf f}l} - \langle S^z_{{\bf f}l} \rangle
\end{equation}
is the fluctuation operator of the $z$-component of the spin and

\begin{equation}
\label{FE}
\langle S^z_{{\bf f}l} \rangle = {\rm Tr} [ S^z_{{\bf f}l}
e^{\beta (F-H-H^\prime(\beta))} ], \qquad \beta=(k_BT)^{-1}
\end{equation}
\noindent
and $F$ is the free energy.

According to the rules of the thermodynamic perturbation expansion
we can write

\begin{equation}
\langle S^z_{{\bf f}l} \rangle = \langle S^z_{{\bf f}l}e^{-\beta H_1}
\rangle_0/\langle e^{-\beta H_1} \rangle_0 , 
\end{equation}
\noindent
where

\begin{equation}
\langle \dots \rangle_0 = {\rm Tr} [ \dots d_0 ]
\end{equation}
and

\begin{equation}
d_0=\{ {\rm Tr} [ \exp(-\beta (H_0+H^\prime(\beta)))]\}^{-1}
\exp(-\beta (H_0+H^\prime(\beta))).
\end{equation}
The right-hand side of expression (\ref{FE})
can be expanded into a series with respect to the perturbing term $H_1$.
From such an infinite series we now choose a certain partial sum
which can be represented graphically in the following way\cite{zo1991}
\vskip -1cm
\begin{equation}
\langle S^z_{{\bf f}l} \rangle = \begin{picture}(30,40)
\put(0,-5){\framebox(36,15)}
\put(6,3){\circle{4}}
\linethickness{3pt}
\put(15,3){\line(1,0){12}}
\put(15,-18){{\bf f}\it l}
\end{picture} \quad,
\end{equation}

\noindent
where

\begin{equation}
\label{obr1}
\begin{picture}(30,30)
\put(0,-5){\framebox(26,15)}
\linethickness{3pt}
\put(6,3){\line(1,0){12}}
\put(9,-18){{\bf f}\it l}
\end{picture}
=
\begin{picture}(30,30)
\put(0,-5){\framebox(26,15)}
\put(12,-18){{\bf f}\it l}
\end{picture}
+
\begin{picture}(30,30)
\put(0,-5){\framebox(30,15)}
\put(6,3){\circle{4}}
\put(22,3){\circle{4}}
\put(8,3){\line(1,0){12}}
\put(13,-18){{\bf f}\it l}
\end{picture}
+
\begin{picture}(52,30)
\put(0,-5){\framebox(52,15)}
\put(5,3){\circle{4}}
\put(21,3){\circle{4}}
\put(30,3){\circle{4}}
\put(46,3){\circle{4}}
\put(7,3){\line(1,0){12}}
\put(32,3){\line(1,0){12}}
\put(24,-18){{\bf f}\it l}
\end{picture}
+
\dots
\end{equation}
\vskip 0.3cm
\noindent
and symbols \vbox{\vskip -0.2cm
\hbox{\textifsym{MM}}}
denote the renormalized interaction line,

\vskip -.2cm
\begin{equation}
\label{obr2}
\begin{picture}(30,30)
\put(7,3){\line(1,0){25}}
\put(-3,-18){{\bf f}\it l}
\put(28,-18){{\bf f}\it l}
\end{picture}
\quad = \quad
\sum_{{\bf f^\prime}l^\prime}
\left(
\dots
\begin{picture}(40,30) \put(0,-5){\framebox(36,15)}
\put(6,3){\circle{4}}
\linethickness{3pt}
\put(11,3){\line(1,0){12}}
\put(28,3){\circle{4}}
\put(-3,-18){{\bf f}\it l}
\put(17,-18){{\bf f$^\prime$}\it l$^\prime$}
\put(36,-18){{\bf f}\it l}
\end{picture}
\dots
\right) ,
\end{equation}

\vskip 0.2cm
\begin{equation}
\begin{picture}(62,30)
\put(0,-5){\framebox(52,15)}
\put(7,3){\circle{4}}
\put(15,3){\circle{4}}
\put(24,2){\dots}
\put(43,3){\circle{4}}
\put(0,-15){$\underbrace {\quad\quad\quad\quad\quad    }$}
\put(20,-30){\it m}
\end{picture}
=
\frac{d^m}{dW^m_{{\bf f}l}}
L(W_{{\bf f}l}) ,
\end{equation}

\begin{equation}
\begin{picture}(40,30)
\put(-8,-10){{\bf f}\it l}
\put(5,2){\dots}
\put(22,-10){${\bf f^\prime}\it l^\prime$}
\end{picture}
= \beta I \sum_{{\bf f^\prime}l^\prime} ,
\end{equation}

\begin{equation}
L(W_{{\bf f}l}) = \ln (2 \cosh(W_{{\bf f}l}/2)) ,
\end{equation}

\begin{equation}
W_{{\bf f}l} = \frac{\beta}{2} \left( I \sum_{{\bf f^\prime}l^\prime}
\langle S^z_{{\bf f^\prime}l^\prime} \rangle
+ I_2 \delta_{l,4} \langle S^z_{r=1} \rangle \right) ,
\end{equation}
\noindent
where we assumed $\hbar = 1$ and $\sum_{{\bf f^\prime}l^\prime}$
denotes summation over nearest neighbours of the ${\bf f}l$ spin.

\begin{widetext}
As a result of calculating the infinite sum (\ref{obr1})
and (\ref{obr2}) we obtain

\begin{equation}
\label{eq22}
\langle S^z_{{\bf f}l} \rangle
= {1\over {2\sqrt{2\pi}}} 
\int_{-\infty}^{+\infty}
e^{-{u^2/2}} \tanh (W_{{\bf f}l} + u \delta W_{{\bf f}l} )du,
\end{equation}
where

\begin{equation}
\delta W_{{\bf f}l} = \frac{\beta}{2}
\left( \frac{1}{2\sqrt{2\pi}}
\int_{-\infty}^{+\infty}
e^{-{u^2/2}} \left[ I^2
\sum_{{\bf f^\prime}l^\prime}
( 1  -\tanh^2(W_{{\bf f^\prime}l^\prime} + u \delta W_{{\bf f}l}))
+  I_2^2 \delta_{l,4} (1-\tanh^2 (V_r + u\delta V_r))\right]
du \right)^{1/2} .
\end{equation}
\end{widetext}

Similarly, using eq. (\ref{Szr1}) for $r=1, 2, \dots$,
the molecular field $V_r$ of the bulk substrate
in the GFA satisfies the equation

\begin{equation}
V_r = \frac{\beta I_1}{2}
\left(\sum_{\bf g} \langle S^z_{{\bf g}r} \rangle + \sum_{{\bf g}r^\prime}
\langle S^z_{r^\prime} \rangle \right) ,
\end{equation}
where

\begin{equation}
\langle S^z_r \rangle = 
\frac{1}{2\sqrt{2\pi}}
\int_{-\infty}^{+\infty}
e^{-u^2/2} \tanh(V_r + u \delta V_r) du ,
\end{equation}

\begin{widetext}
\begin{equation}
\label{eq26}
\begin{split}
\delta V_r = \frac{\beta I_1}{2}
( \frac{1}{\sqrt{2\pi}}
\int_{-\infty}^{+\infty}
e^{-u^2/2} & [ 4 (1-\tanh^2 (V_r+u\delta V_r))\\
& + (1-\delta_{r,1})(1-\tanh^2(V_{r-1}+u\delta V_{r-1}))
 + (1-\tanh^2 (V_{r+1} + u\delta V_{r+1})) ] du)^{1/2}
\end{split}
\end{equation}
and $\delta W_{{\bf f}l}$, $\delta V_r$ are the mean
Gaussian fluctuations of the molecular fields $W_{{\bf f}l}$
and $V_r$, respectively.

After introducing the following reduced magnitudes,

\begin{equation}
X_{{\bf f}l} = 2 \langle S^z_{{\bf f}l} \rangle , y_r=2 \langle S^z_r \rangle_0,\\
t=\frac{4}{\beta I} ,  a=\frac{I_2}{I} , b=\frac{I_1}{I} ,
\end{equation}
equations (24) 
-(\ref{eq26}) can be written in a compact form,

\begin{equation}
\label{eq28}
X_{{\bf f}l}=
\frac{1}{\sqrt{2\pi}}
\int_{-\infty}^{+\infty}
e^{-{u^2/2}} \tanh (W_{{\bf f}l} + u \delta W_{{\bf f}l} )du,
\end{equation}
where

\begin{equation}
W_{{\bf f}l} = \frac{1}{t}
\left(
\sum_{{\bf f^\prime}l^\prime}
X_{{\bf f^\prime}l^\prime} +
a\delta_{l,4} y_{r=1} \right) ,
\end{equation}

\begin{equation}
\begin{split}
\delta W_{{\bf f}l} =
{2\over t} (
\frac{1}{\sqrt{2\pi}}
\int_{-\infty}^{+\infty}
e^{-u^2/2} [
& 1  -\tanh^2 (W_{{\bf f}l} + u\delta W_{{\bf f}l} )
+ \sum_{{\bf f^\prime}l^\prime}
(1-\tanh^2(W_{{\bf f^\prime}l^\prime}
+ u \delta W_{{\bf f^\prime}l^\prime}))\\
& +  a^2 \delta_{l,4}
(1-\tanh^2 (V_r + u\delta V_r)) ] du )^{1/2} ,
\end{split}
\end{equation}

\begin{equation}
V_r=\frac{b}{t} (4y_r + y_{r+1} + (1-\delta_{r,1})y_{r-1}) ,
\end{equation}

\begin{equation}
y_r=
\frac{1}{\sqrt{2\pi}}
\int_{-\infty}^{+\infty}
e^{-u^2/2} \tanh (V_r + u\delta V_r) du ,
\end{equation}

\begin{equation}
\begin{split}
\delta V_r =
\frac{2 b}{t}
(
\frac{1}{\sqrt{2\pi}}
\int_{-\infty}^{+\infty}
e^{-u^2/2} [
& 4  (1-\tanh^2(V_r+u\delta V_r))
+  (1-\delta_{r,1})(1-\tanh^2(V_{r-1}+u\delta V_{r-1}))\\
& + (1-\tanh^2(V_{r+1} + u\delta V_{r+1}) ]
du )^{1/2} \quad .
\end{split}
\end{equation}
\end{widetext}

Equations (30)-(35)
were solved numerically. In order to carry out
the calculations we have to specify the magnetic lattices pyramid
and substrate as a simple cubic one. Results are presented graphically.

Fig. 2 shows the spatial distributions of magnetization
$X_4$ in the $l=4$ layer of the nanoscopic pyramid for
temperature $t=4.0$ and $a=0.5$, $b=1.0$ in both
MFA and GFA.

Figs. 3-5 present spatial distributions of Gaussian fluctuations
of molecular field $\delta W_{{\bf f}l}$ for $l=2, 3, 4$
respectively, temperature $t=4.0$ and $a=0.5$, $b=1.0$.
In the tip layer adjacent to the substrate the magnitude
of fluctuations is largest in the second row of atoms,
counting from the outside. Comparing Fig. 4 and 5 we see that
maximum fluctuations occur along the edges of the pyramid.
Smaller fluctuations of magnetic moments in the center of the base
of the apex is an effect of the substrate.

Fig. 6 shows the mean value of magnetization

\begin{equation}
X_l = {{\sum_{\bf f} X_{{\bf f}l}}\over \sum_{\bf f}} \quad ,
\end{equation}
in each monoatomic layer of pyramid
$l=1, 2, 3, 4$ in GFA as a function of temperature
$t$ for $a=0$, $b=1$ and $a=0.5$, $b=1.0$.

Fig. 7 presents temperature dependence of the mean pyramid
magnetization

\begin{equation}
X=(4X_1+16X_2+36X_3+64X_4)/120
\end{equation}
obtained in GFA for $a=0$ and $a=0.5$ and $b=0.5, 1.0, 2.0$.
Fig. 8 shows magnetization of individual layers of the pyramid
and the substrate. The influence of the pyramid extends several layers
into the substrate. Similar conclusion was reached
in an earlier study\cite{gautier1995} of the interaction
between a magnetic nanotip and a magnetic surface using the tight-binding
model. The tight-binding calculation for a Fe tip on a Fe surface
shows that only the first four layers of the tip support
are affected. The inclusion of fluctuations in our work
increase the depth of this influence as $t \rightarrow t_c$, see Fig. 8.

Fig. 9 presents the mean Gaussian fluctuation of molecular field

\begin{equation}
\delta W_l = \frac{\sum_{{\bf f}l} \delta W_{{\bf f}l}}
{\sum_{{\bf f}l}}
\end{equation}
and $\delta V_r$ for pyramid and the substrate respectively.

Finally, Fig. 10 shows Gaussian fluctuation of the molecular field
as a function of temperature.

\begin{table}
\caption{\label{table1}Curie temperature $t_c$ (in relative units)
for the nanoscopic pyramid.}
\begin{ruledtabular}
\begin{tabular}{cccc}
& &\multicolumn{2}{c}{$t_c$}\\
$a$&$b$&{\rm MFA}&{\rm GFA}\\
\hline
0&arbitrary&~4.93&3.86\\
0.1&0.5&~4.93&1.88\\
0.1&1.0&~5.99&4.13\\
0.1&2.0&11.95&9.30\\
0.5&1.0&~5.99&4.13\\
0.5&2.0&11.95&9.27\\
\end{tabular}
\end{ruledtabular}
\end{table}

In Table 1 the Curie temperature $t_c$ of the ferromagnetic nanoscopic
pyramid is given in relative units 
for different values of parameters $a$ and $b$ obtained
in MFA and GFA.

We can see that for a nanopyramid the ratio of the Curie
temperature in GFA to that obtained in MFA is $0.783$.
For an infinite simple cubic lattice
this ratio is equal to $0.856$ and for a square lattice
Ising monolayer it has a value of $0.799$.\cite{zo1980A}
The ratio of the Curie temperature of a monolayer
obtained in GFA to the exact result is $1.409$.
The same quantity for the sc bulk Ising ferromagnet
is approximately $1.138$.


\section{Conclusions}

The results we obtained lead us to the following conclusions:

\begin{itemize}
\item{The interation of a ferromagnetic nanoscopic pyramid
with its bulk ferromagnetic substrate may have essential influence
on the properties of the ferromagnetic pyramid,}

\item{The distribution of Gaussian fluctuations is highly
nonuniform,
}

\item{$\lim_{t\rightarrow 0} \delta W_{\textbf{f}l} = 0$,
as expected,}

\item{The maximum of molecular field fluctuations
$\delta W_{\textbf{f}l}$ is reached at the Curie
temperature $t_c$, as expected,
}

\item{The Curie temperature $t_c$ of the pyramid strongly depends
on the Curie temperature of the bulk substrate,}

\item{We believe that ERSO (\ref{rdo})
may be successfully applied in studies of the influence
of the bulk substrate on the thermodynamic properties of nonmagnetic
systems and other more complex nanoscopic
systems.\cite{giersig,giersig2}}
\end{itemize}

The development of the magnetic exchange force microscopy
(MExFM)\cite{wiesendanger2007} may enable studies of the spatial
distribution of magnetization in nanopyramides deposited
on nonmagnetic surface. Such magnetic particles are of interest
in magnetic recording techniques.

It would be useful to study the temperature dependence
of exchange force between the nanotip
and a well-characterized magnetic surface
of a material with Curie temperature higher than $T_c$ of the tip.
On cooling the system one should observe the magnetic
transition of the tip. The force of the interaction
between the tip and the surface is measured as the shift
of frequency of oscillations of the cantilever. As $T$ decreases,
magnetization of the tip increases and so does the exchange
interaction with the surface, resulting in a frequency shift.
If the ratio of exchange energies $I_1/I_2$ is large, i.e. when exchange
interaction between atoms within the tip is much larger than exchange
between the bottom layer of the tip and the first layer of
the substrate, or when the substrate is nonmagnetic,
fluctuations may lead to a long tail of magnetization
as a function of temperature, see Fig. 7.
In this case magnetization may saturate at temperatures much smaller
than $T_c$.

The measurement of the frequency
shift as a function of temperature at fixed position
above the surface may provide a good estimate
of possible magnetic fluctuation effects.
The crossover temperature 
between the tail, see $b=2$ curve
in Fig. 7, and the saturated part at low $T$
detected via the oscillation frequency shift
should not depend on position above the surface.
Temperature scans at different
surface positions would result in frequency
shifts of different magnitudes occurring at approximately
the same temperature.

Recent attempts to map single spins on the surface
of antiferromagnetic insulator NiO were only partially
successful. The experiment
of Kaiser et al.\cite{wiesendanger2007} with Fe tip
performed  at about 10K in a magnetic field of 5 T showed clear picture
of antiferromagnetic structure of the (001) surface of nickel oxide.
It cannot be ruled, however, that the applied magnetic field
induces structural changes in NiO.
The measurements of Schmid et al.\cite{mannhart2008}
conducted at room temperature, using Co and NiO tips did not
did not show any spin contrast.
The ordering temperature of NiO is 525 K. The authors point out
that the magnetic ordering
temperature of the tip may be significantly reduced
relative to the bulk value. Our calculation shows that
it is also possible that the spin ordering of the tip is present
but is very weak. Performing this experiment at lower temperatures,
as authors of Ref. \cite{mannhart2008} suggest,
may yield better results.

Atomically sharp magnetic tips are crucial in achieving
lateral high spatial resolution. Therefore
the knowledge of spatial distribution of fluctuations
and its temperature dependence might improve
the calibration of the STM device and help in estimating
experimental error of the MExFM method.
For small distances between the tip and the surface
it may be necessary to include interaction
of more than one atom of the tip or of the surface.\cite{baratoff2003}
In this case our model with nonuniform magnetization
of the tip may provide useful insight. The temperature dependence
of mean value of magnetization
of the individual tip layers shown in Fig. 6 shows that
the magnetization in the second layer can be significantly larger
that the magnetization of the apex. This implies
an increased range of tip-surface
separation where
the interaction with atoms in the next layer
should be taken into account.


The ability to control magnetic properties of nanoclusters\cite{heinrich2006}
is one of the central problems of nanotechnology.
Arrays of nanoclusters are also intensively
studied.\cite{mueller2001,barski2006}
The theory presented here may be applied
to various geometrical shapes used in experiments,\cite{skomski2003} e.g.
chains of particles, striped and cylindrical nanowires,
nanodots, nanojunctions, surface steps.
We intend to carry out calculations in a more realistic
model of the substrate surface and the interface
between the pyramid and the substrate.

Finally let us note that ERSO (\ref{rdo})
couples the nanoscopic pyramid with the substrate.
This means that the system does not have a finite number
of spins. Therefore the use of e.g. Monte Carlo simulations
would have little justification.

\acknowledgments
Some of the computations were performed in the Computer Center
of the Tri-city Academic Computer Network in Gdansk.

\vfil
\eject

\begin{figure}[1htb!]
\includegraphics[width=\columnwidth,clip]{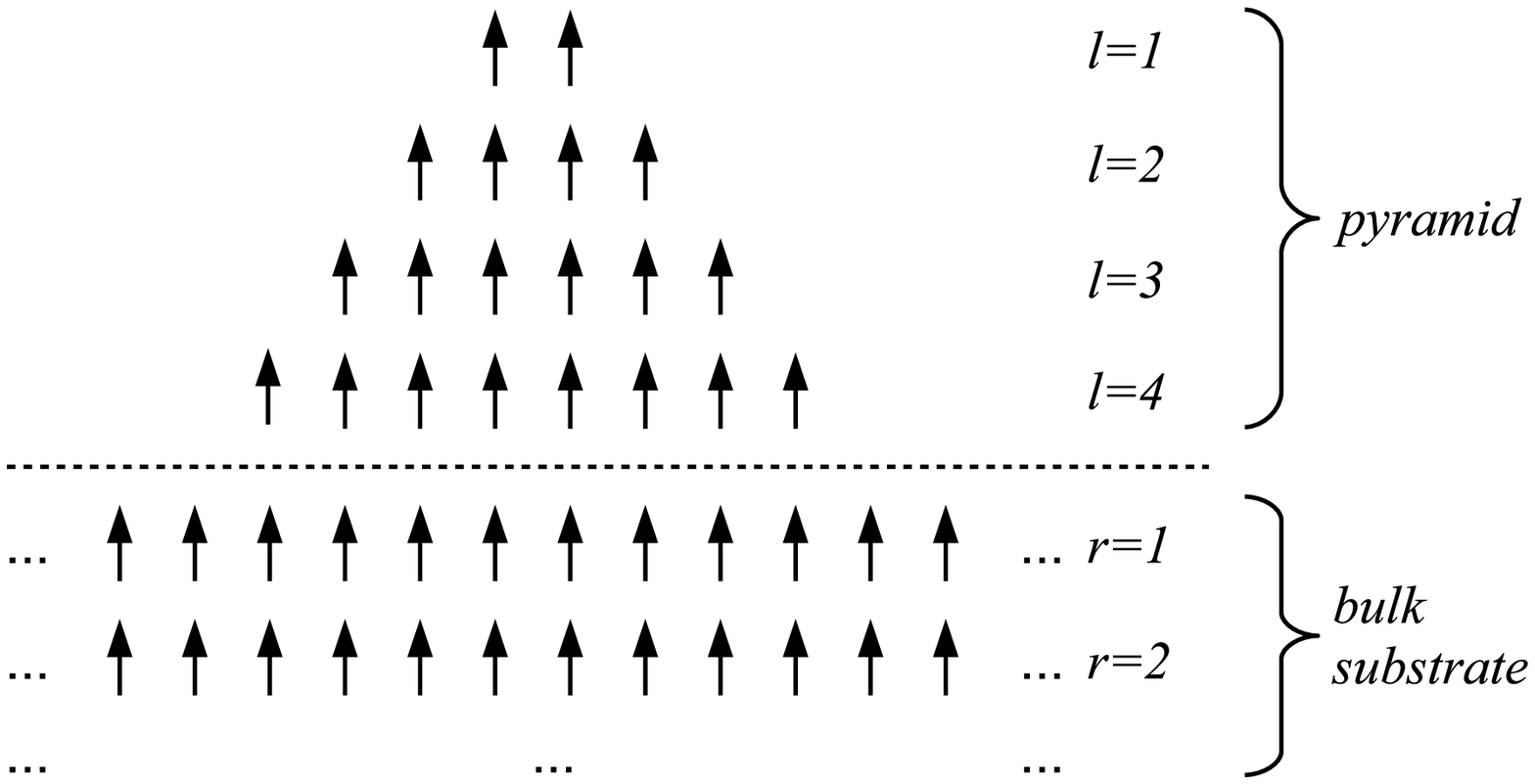}
\caption{
Schematic diagram of ferromagnetic nanoscopic pyramid deposited
on a ferromagnetic bulk substrate.}
\label{fig1}
\end{figure}

\begin{figure}[2htb!]
\includegraphics[width=0.9\columnwidth]{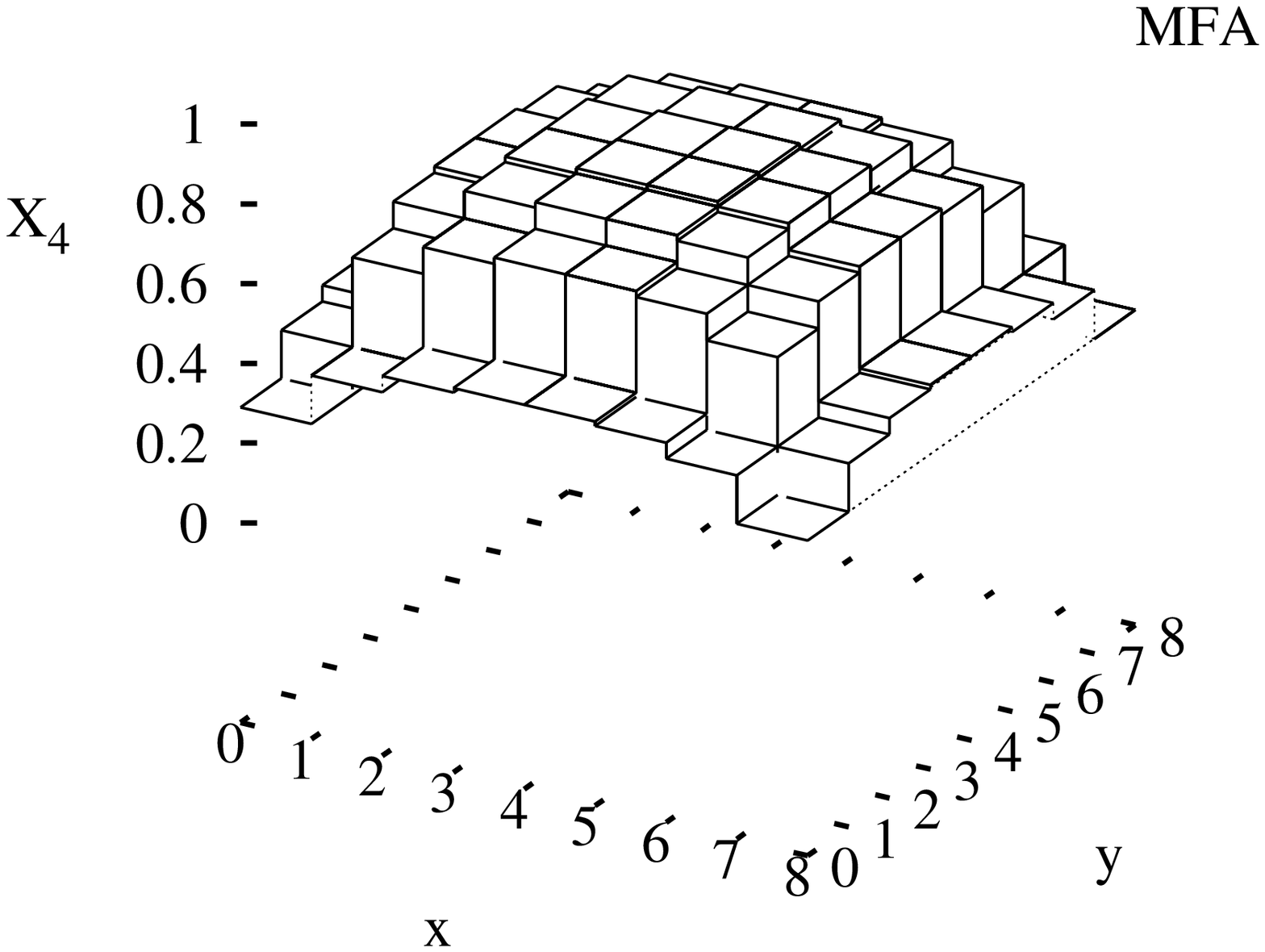}
\vskip -2cm
\includegraphics[width=0.9\columnwidth]{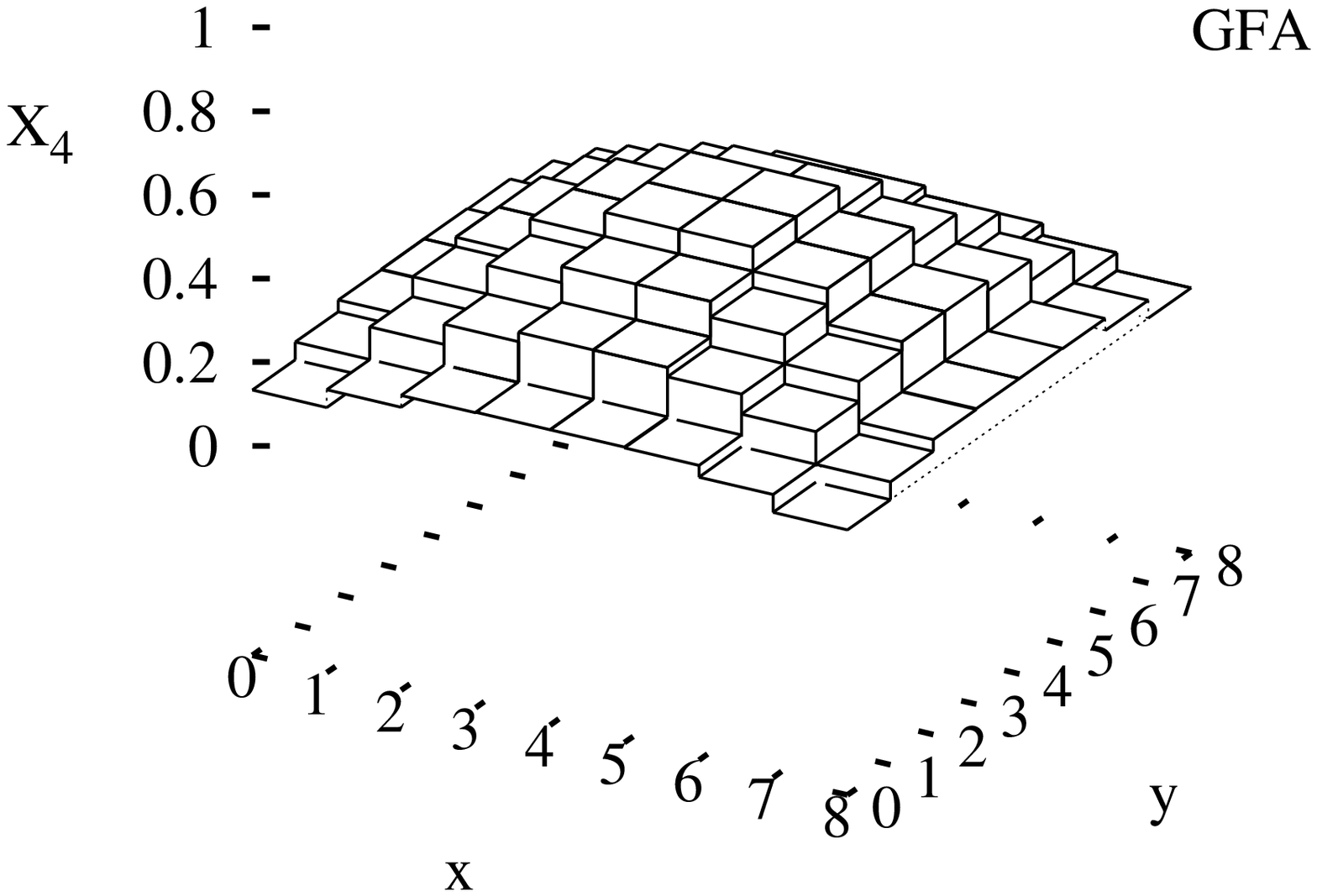}
\vskip -2.5cm
\caption{
Spatial distribution of magnetization in nanoscopic pyramid for
$l=4$, temperature $t=4.0$, and $a=0.5$, $b=1$ in MFA\cite{lsb2007}
and GFA.
}
\end{figure}

\begin{figure}[3ht!]
\includegraphics[width=0.9\columnwidth]{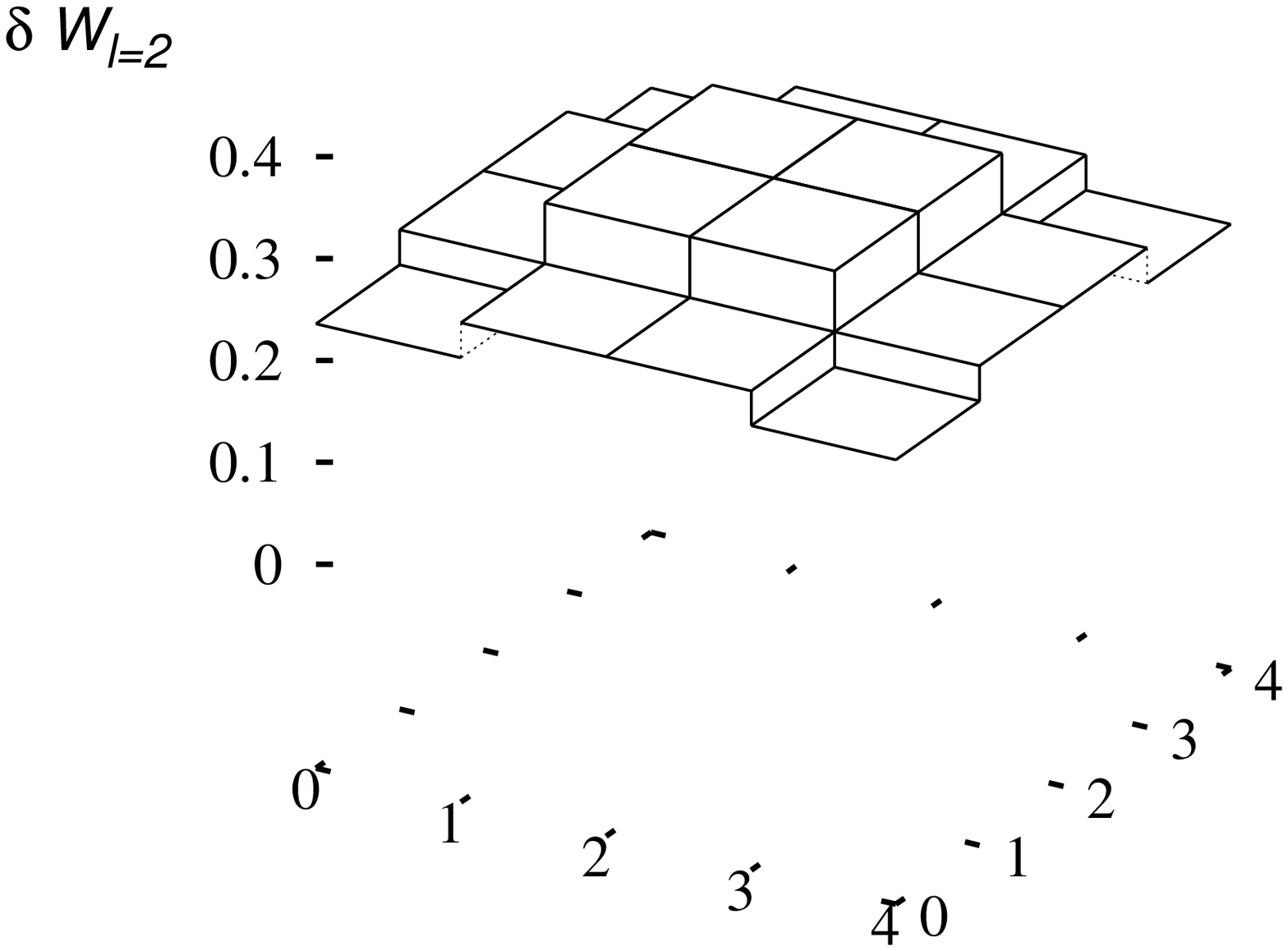}
\vskip -2.5cm
\caption{
Spatial distribution of Gaussian fluctuations
of the molecular field in pyramid layer 
$l=2$ at temperature $t=4.0$ and $a=0.5$, $b=1$. 
}
\end{figure}

\begin{figure}[4ht!]
\includegraphics[width=0.9\columnwidth]{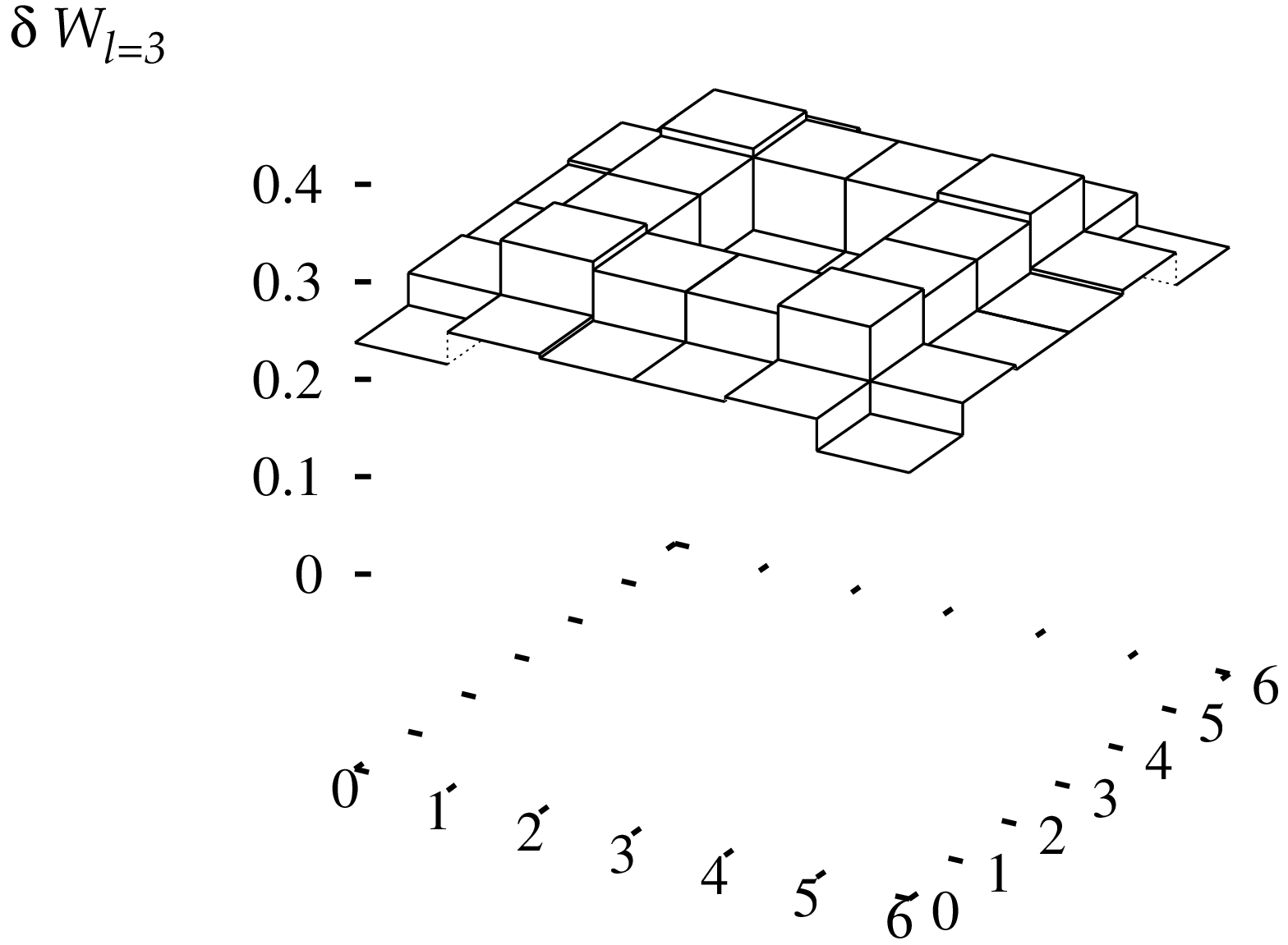}
\vskip -2.5cm
\caption{
Spatial distribution of Gaussian fluctuations
of the molecular field in pyramid layer
$l=3$ at temperature $t=4.0$ and $a=0.5$, $b=1$.
}
\end{figure}

\begin{figure}[5htb!]
\includegraphics[width=0.9\columnwidth]{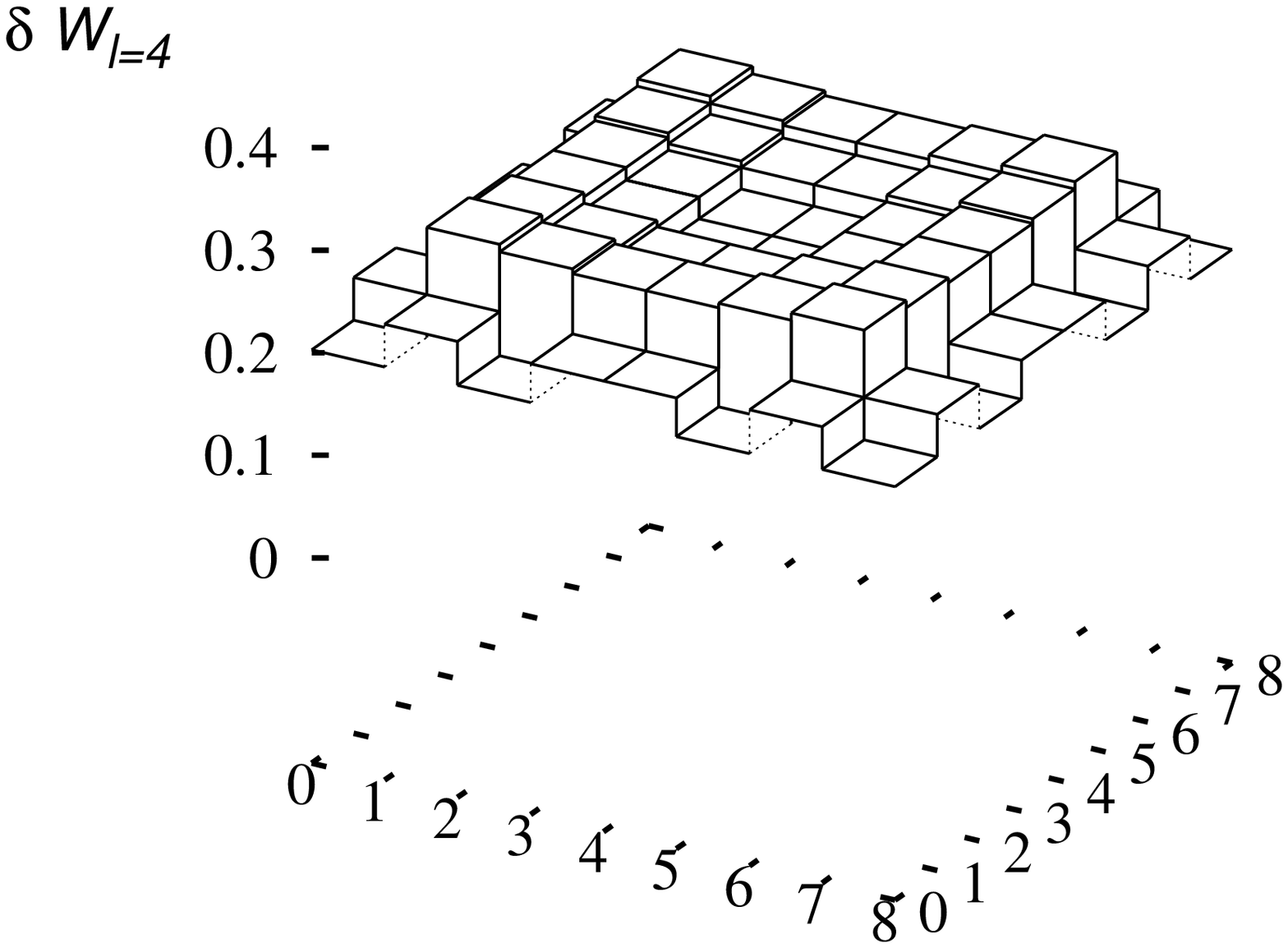}
\vskip -2.5cm
\caption{
Spatial distribution of Gaussian fluctuations
of the molecular field in pyramid layer
$l=4$ at temperature $t=4.0$ and $a=0.5$, $b=1$.
}
\end{figure}

\begin{figure}[6ht!]
\includegraphics[width=0.7\columnwidth]{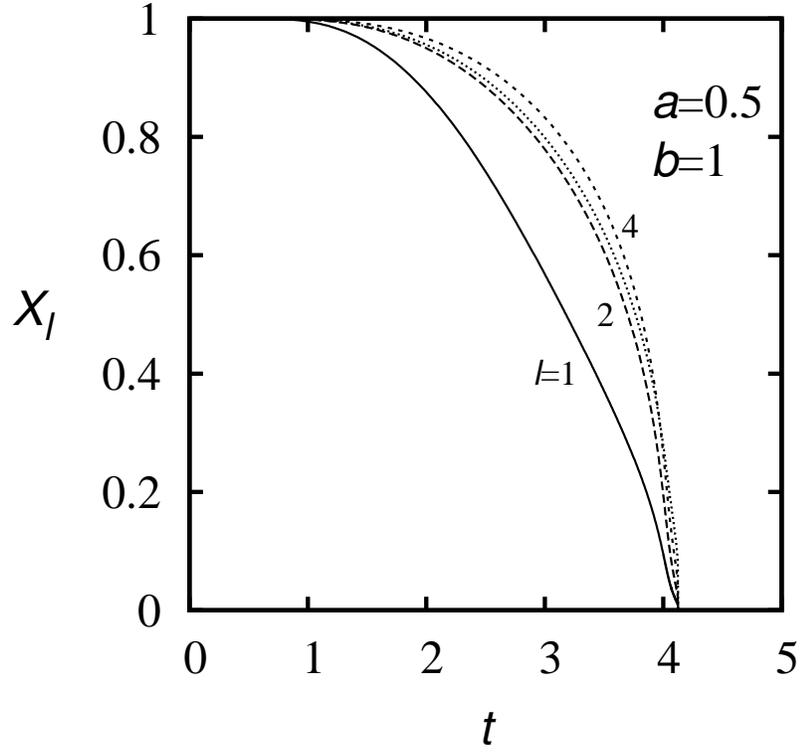}
\vskip -1.9cm
\caption{
Mean value of magnetization,
$X_l = \sum_{{\bf f}l} X_{{\bf f}l}/\sum_{\bf f}$
in each monoatomic layer of pyramid, $l=1, 2, 3, 4$,
in GFA as a function of temperature $t$ for
$a=0.5$, $b=1.0$.
}
\end{figure}

\begin{figure}[7ht!]
\includegraphics[width=0.7\columnwidth]{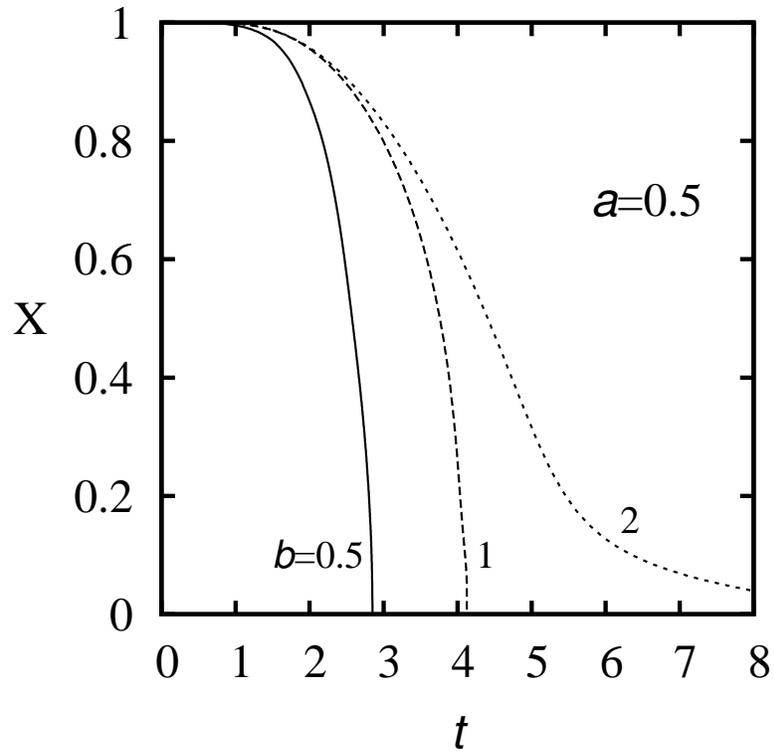}
\vskip -2.cm
\caption{
Temperature dependence
of the mean pyramid magnetization
${\bf X} = (4X_1 + 16X_2 + 36X_3 + 64X_4)/120$ for
$a=0.5$ and $b=0.5, 1, 2$.
}
\end{figure}

\vskip -0.3cm
{\begin{figure}[9htb!]
\includegraphics[width=0.8\columnwidth]{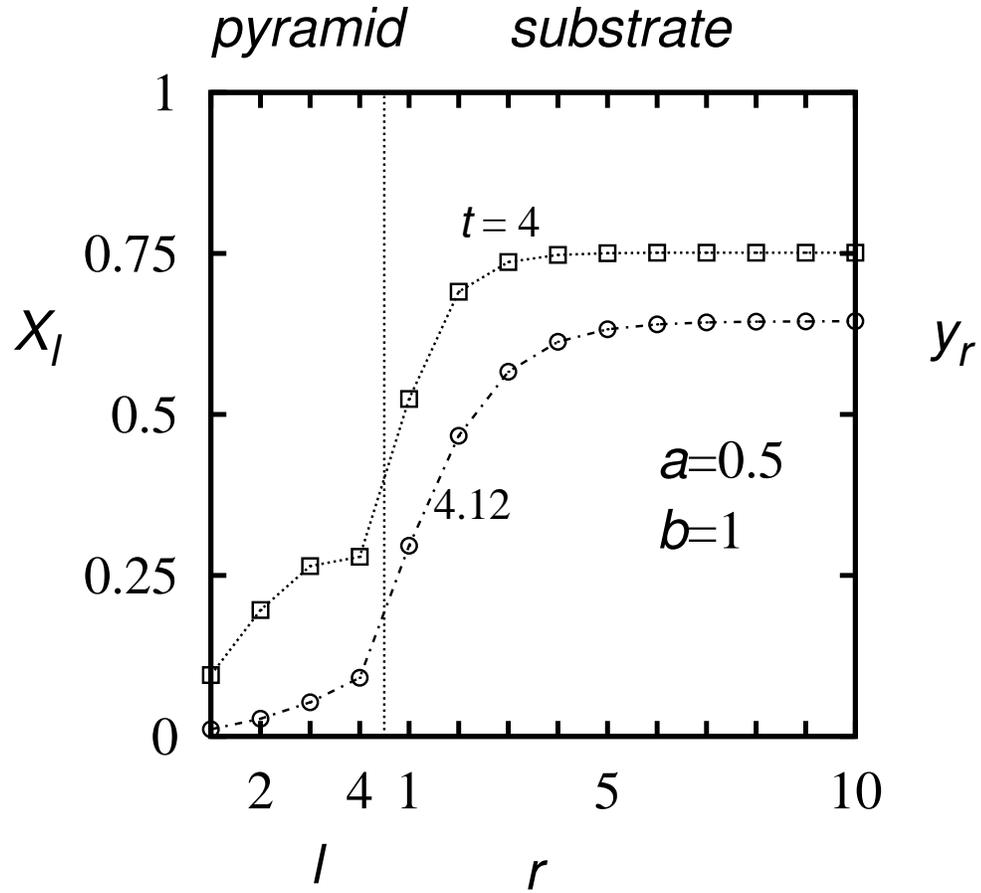}
\vskip -2.2cm
\caption{
Magnetization in individual layers of the pyramid $X_l$
and the substrate $y_r$ for $a=0.5$, $b=1.0$ and $t=4.0, 4.12$}
\end{figure}
}

\begin{figure}[10htb!]
\includegraphics[width=0.8\columnwidth]{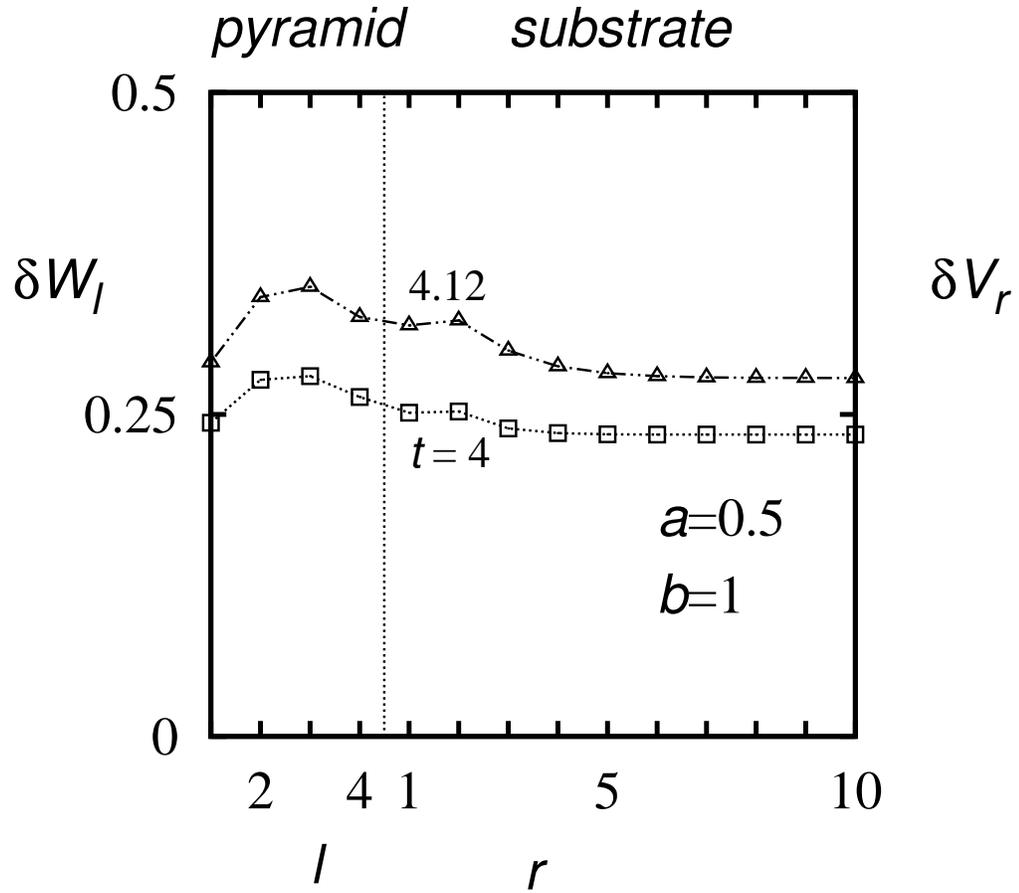}
\vskip -2.2cm
\caption{
Gaussian fluctuation of the molecular field
in individual layers of the pyramid
and the substrate for $a=0.5$, $b=1.0$ and $t=4.0, 4.12$.
}
\end{figure}

\begin{figure}[11ht!]
\includegraphics[width=0.8\columnwidth]{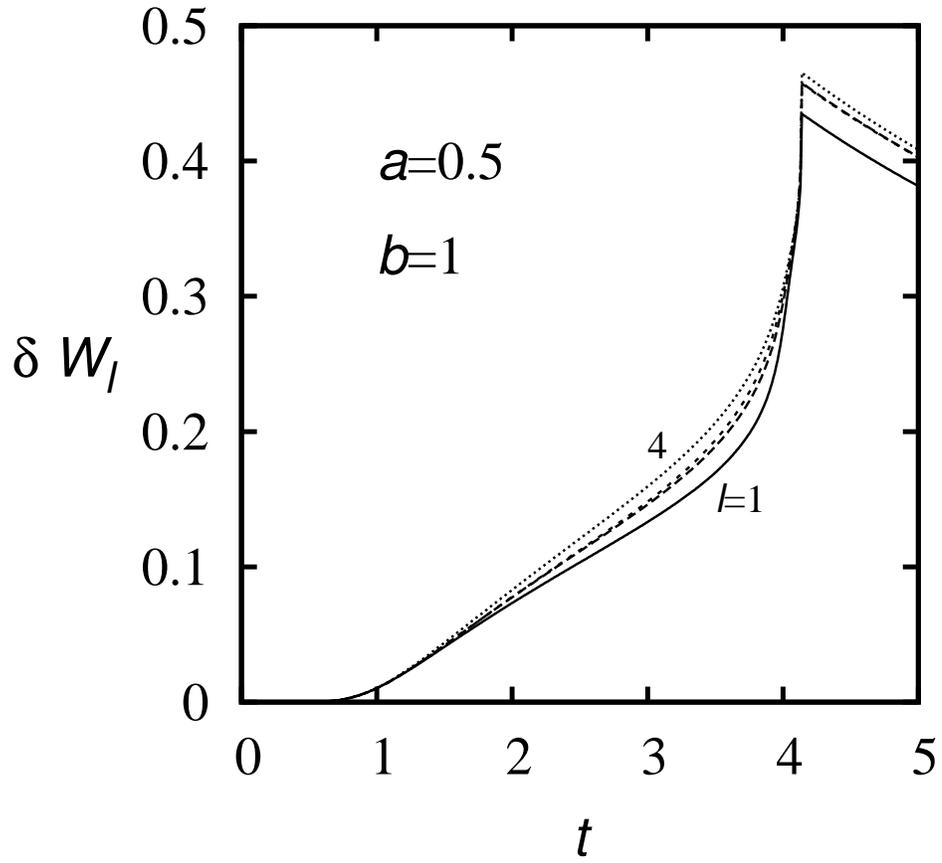}
\vskip -2.2cm
\caption{
Temperature dependence of Gaussian fluctuations of molecular field
in individual layers of the pyramid $\delta W_l$
for $a=0.5$ and $b=1.0$.
}
\end{figure}

\end{document}